\documentclass[aps,prl,twocolumn,superscriptaddress,showpacs]{revtex4}
\usepackage{graphicx}
\usepackage{amsmath}
\usepackage{natbib}
\begin{document}
\title{Scarring by homoclinic and heteroclinic orbits}
\author{D. A. Wisniacki}
\affiliation{Departamento de F\'\i sica ``J. J. Giambiagi'', FCEN,
UBA, Pabell\'on 1, Ciudad Universitaria, 1428 Buenos Aires, Argentina.}
\author{E. Vergini}

\affiliation{Departamento de F\'{\i}sica,
 Comisi\'on Nacional de Energ\'{\i}a At\'omica.
 Av.\ del Libertador 8250, 1429 Buenos Aires, Argentina.}
\author{R. M. Benito}
\affiliation{Grupo de Sistemas Complejos
 and Departamento de F\'{\i}sica,
 Escuela T\'ecnica Superior de Ingenieros
 Agr\'onomos, Universidad Polit\'ecnica de Madrid,
 28040 Madrid, Spain.}
\author{F. Borondo}
\affiliation{Departamento de Qu\'{\i}mica C--IX,
 Universidad Aut\'onoma de Madrid,
 Cantoblanco, 28049--Madrid, Spain.}
\date{\today}
\begin{abstract}
In addition to the well known scarring effect of periodic orbits,
we show here that homoclinic and heteroclinic orbits,
which are cornerstones in the theory of classical chaos,
also scar eigenfunctions of classically chaotic systems
when associated closed circuits in phase space are properly quantized,
thus introducing strong quantum correlations.
The corresponding quantization rules are also established.
This opens the door for developing computationally tractable methods
to calculate eigenstates of chaotic systems.
\end{abstract}
\maketitle
%

Scarring constitutes an important topic in the field of quantum chaos
\cite{Heller1,Bogomolny,Berry2,Ozorio,Tomsovic,Kaplan,Keating},
representing a dramatic departure from random matrix theory predictions \cite{Kaplan},
and providing many practical applications \cite{RTD}.
The term ``scar'' was introduced by Heller to describe an enhanced probability
density in certain eigenfunctions of classically chaotic systems taking place
along short periodic orbits (POs) \cite{Heller1,Heller2}.
In order to explain this phenomenon, he noted that recurrences along a PO can
compete under suitable conditions with the unstable dynamics in its vicinity,
giving rise to the building up of probability.
From this point of view, an interesting speculation \cite{Ozorio,speculation}
concerns the possibility that long time motions, such as homoclinic or
heteroclinic orbits, which are cornerstones in the classical theory of chaos,
could also generate scars.

In this Letter, we demonstrate that such motions can indeed produce
enhanced probability density accumulation associated with POs.
However, this new type of localization differs in two aspects
from that previously described in the literature \cite{Heller1}.
In the case of scars produced by POs, quantized closed
circuits are defined in the direction of the dynamical flux,
and attention is focussed into the part of the probability
remaining near the PO after recurrences.
In our case, the relevant circuits to be quantized are defined in a
transverse direction to the flux, clearly observed on a surface of section,
and the attention is concentrated on the probability leaving and then
returning to the vicinity of the PO.
This is a nontrivial fact, since it implies that the time scales
characterizing these processes are of very different nature.
While PO scarring takes place in times of the order of the PO period,
the localization induced by homoclinic and heteroclinic excursions needs
a time of the order of the Ehrenfest time, $t_E$, to develop.

The model used in our calculations is the fully chaotic system consisting of a
particle confined in a desymmetrized Bunimovich stadium billiard, defined by
the radius of the circular part, $r=1$, and the enclosed area, $1+\pi/4$.
In the quantum calculations, Dirichlet conditions on the stadium boundary and
Neumman conditions on the symmetry axes are imposed.
%
\begin{figure}[t]
 \includegraphics[width=7.0cm]{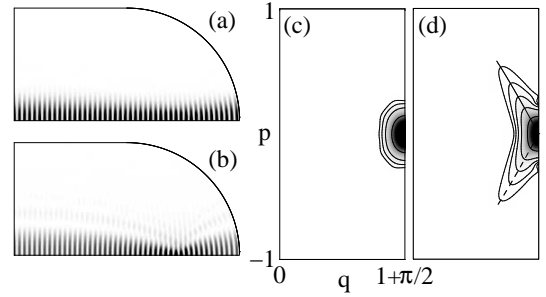}
 \caption{(a) Tube and (b) scar wave functions along the horizontal periodic
 orbit of a desymmetrized stadium billiard corresponding to the quantum
 number $n_H=44$.
 The associated Husimi based quantum surfaces of section, in Birkhoff coordinates,
 are shown in (c) and (d).
 The manifolds of the horizontal periodic orbit are also plotted in panel (d).}
 \label{fig:1}
\end{figure}
%
\begin{figure}
 \includegraphics[width=7.0cm]{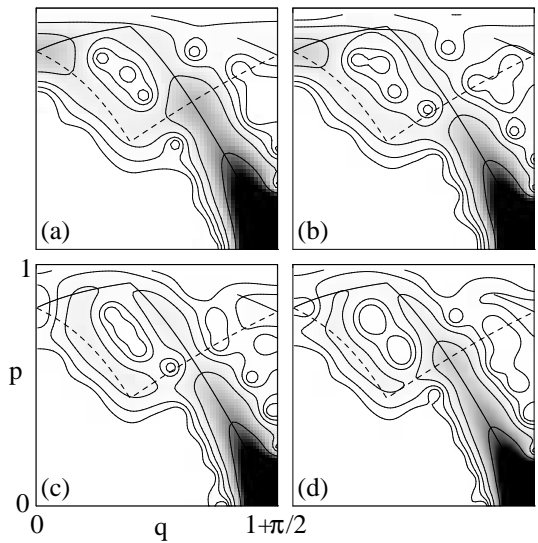}
 \caption{Husimi based quantum surfaces of section for the scar functions
 corresponding to the quantization condition summarized in
 Table~\protect\ref{table:I}, showing the localization effects produced
 by the homoclinic motion on ${ho}_1$ and ${ho}_2$, when the corresponding
 areas (see Fig.~\protect\ref{fig:3}) are quantized.
 To guide the eye we have superimposed the manifolds corresponding
 to the horizontal periodic orbit.}
 \label{fig:2}
\end{figure}
We focus our attention on the dynamics influenced by the horizontal PO.
To study scarring phenomena related to this orbit,
we define scar functions in the following way.
We start from ``tube functions'', $|\phi_{\rm tube} \rangle$,
covering the region around the PO, obtained with the semiclassical theory
of resonances developed in Ref.~\onlinecite{Vergini2}.
Associated to the construction of these wave functions there is a
quantization on the PO energy, $E_{\rm BS}$, given by the usual
Bohr--Sommerfeld (BS) condition on the wavenumber
%
\begin{equation}
  k_{\rm BS}=\frac{2\pi}{L_H} \left( n_H+\frac{\nu_H}{4} \right),
  \label{eq:BS}
\end{equation}
being $n_H$ the quantum number along the orbit, $L_H=4$ its length,
and $\nu_H=3$ the corresponding Maslov index.
This condition guarantees the coherent contribution of probability
recurrences along the PO.
An example of these tube functions is shown in Fig.~\ref{fig:1}(a).
When examined in phase space, using a Husimi based quantum surface of
section (QSOS) \cite{Wisniacki1},
this function appears localized around the associated classical fixed
point at $(q,p)_{(A)}=(1+\pi/2,0)$; see Fig.~\ref{fig:1}(c).
In a second step the dynamical information up to a given time, $T$,
is incorporated into these functions by using the dynamical averaging method
developed in Ref.~\onlinecite{Polavieja}
\begin{equation}
  |\phi_{\rm scar} \rangle =
     \int_{-T}^{T}\! dt\; \cos \! \left(\! \frac{\pi t}{2T}\! \right) \;
     e^{i (E_{\rm BS}-\hat{H}) t/\hbar} \; | \phi_{\rm tube}\rangle,
  \label{eq:scar}
\end{equation}
where the cosine function has been included in order to minimize the energy
dispersion of $|\phi_{\rm scar} \rangle$ \cite{Vergini3}.
In this way, all processes transporting probability from and to the
region defined by $| \phi_{\rm tube}\rangle$, in times smaller than $T$ and with
energies in the window $E_{\rm BS}\pm \hbar/T$, will be taken into account
\cite{note0}.
An example for $T=0.95 t_E$ is shown in Figs.~\ref{fig:1}(b) and (d).
As can be seen, a self--focal point appears in the wave function [Fig.~\ref{fig:1}(b)]
and the phase space density spreads along the hyperbolic structure as the manifolds
go away from the PO [Fig.~\ref{fig:1}(d)].
We emphasize that in this case, the only channel able to transport probability
from and to the region defined by $| \phi_{\rm tube}\rangle$ is the PO,
this being the reason that a semiclassical recipe for the construction of scar
functions can be easily derived \cite{Vergini1}.

%
\begin{table}
  \begin{tabular}{ccccc}
   \hline \hline
   Label   & $n_H$ & $k_{\rm BS}$ & $n_{{ho}_1}$ & $n_{{ho}_2}$ \\
   \hline
   (a)     & 34    & 54.585       & 29.01        & 25.99   \\
   (b)     & 40    & 64.010       & 34.07        & 30.47   \\
   (c)     & 44    & 70.293       & 37.43        & 33.46   \\
   (d)     & 50    & 79.718       & 42.49        & 37.95   \\
   \hline \hline
  \end{tabular}
  \caption{Quantum number and Bohr--Sommerfeld quantized wavelength along
  the horizontal PO [Eq.~(\ref{eq:BS})], and homoclinic quantum numbers
  [Eq.~(\ref{eq:homo})] corresponding to the cases shown in
  Fig.~\protect\ref{fig:2}.}
  \label{table:I}
\end{table}
\begin{figure}[b]
 \includegraphics[width=8.5cm]{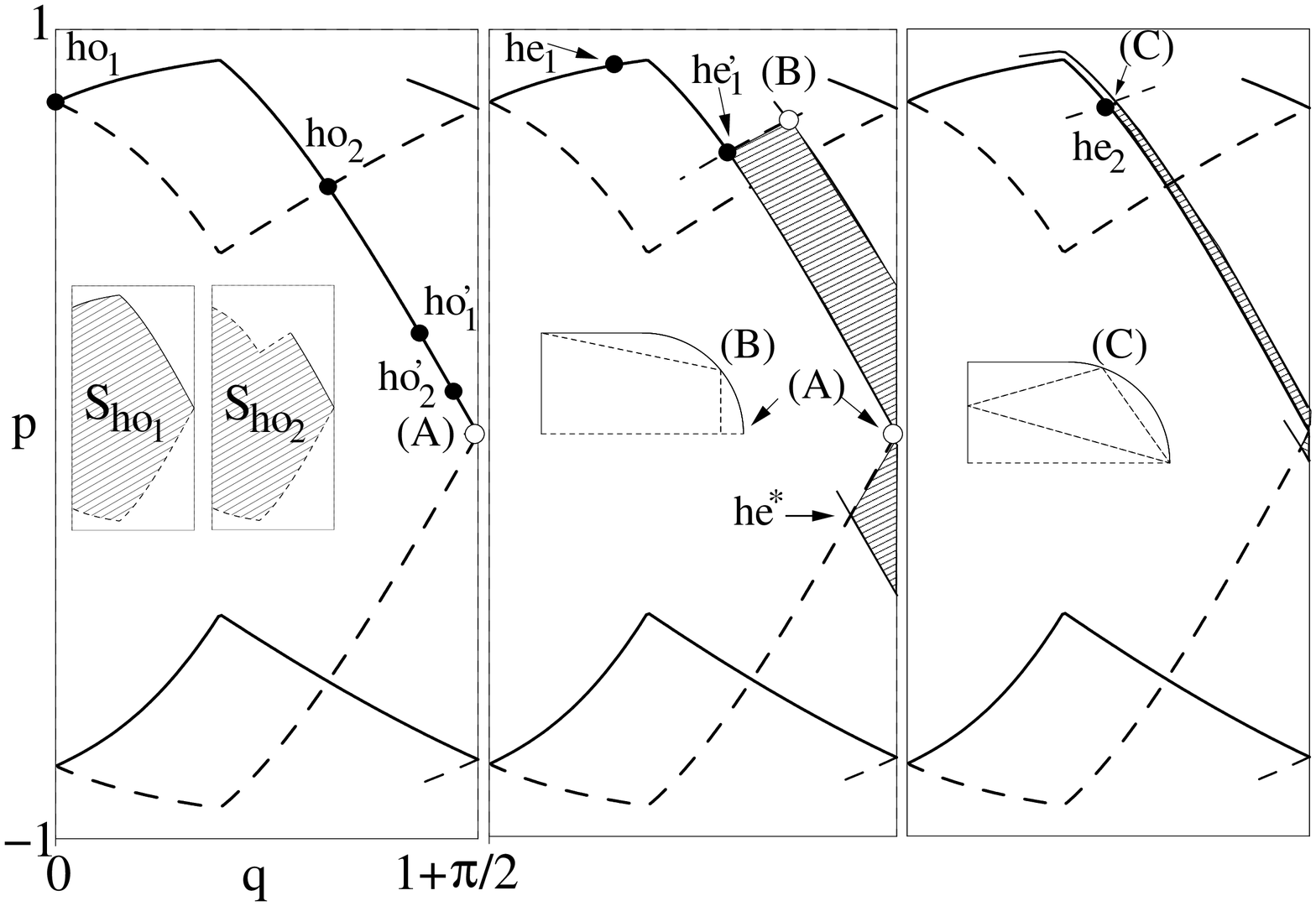}
 \caption{Detail of the phase space portrait for the desymmetrized stadium
  billiard in Birkhoff coordinates. \protect \\
  (Left) Fixed point, (A), and unstable (full line) and stable (dashed line)
  manifolds corresponding to the horizontal periodic orbit.
  These manifolds first cross, with different topology, at the primary
  homoclinic points, ${ho}_1$ and ${ho}_2$, which map (backwards in time) into
  ${ho}'_1$, $\ldots$, and ${ho}'_2$, $\ldots$, respectively.
  They define two primary homoclinic areas, $S_{{ho}_1}$ and $S_{{ho}_2}$,
  which are shown in the inset. \protect \\
  (Center) Fixed point, (B), and manifolds for the
  ``flipped L''--shaped periodic orbit (see inset).
  The heteroclinic points with the horizontal orbit, $he'_1$ and $he^*$,
  and the corresponding heteroclinic area, $S_{{he}_1}$,
  are also shown. \protect \\
  (Right) Same for the ``triangle'' orbit.}
 \label{fig:3}
\end{figure}
Let us now consider longer propagation times, thus opening the possibility
to observe more interesting dynamical effects.
In Fig.~\ref{fig:2} we show QSOS for some scar functions calculated by means
of Eq.~(\ref{eq:scar}) with $T=1.40t_E$, so that the effects of the
primary homoclinic motion associated to the horizontal PO have had time
to fully develop (notice the different range spanned by the vertical axes
with respect to Fig.~\ref{fig:1}).
The different panels correspond to four BS quantization conditions on the
horizontal PO, which are summarized in Table \ref{table:I}.
As can be seen, all QSOS show a prominent peak on the horizontal PO fixed point,
with much of the probability density spreading all along the full length
(linear and nonlinear parts) of the emanating manifolds.
Furthermore, a closer examination reveals conspicuous accumulations of density
on different points along these manifolds.
To analyze in detail which are these points, we present in Fig.~\ref{fig:3}
a detailed phase space portrait of the classical structures related to the
scarring orbit.
In the left part we see the stable and unstable manifolds emanating from
fixed point (A), which first cross (with different topology) at the two primary
homoclinic points $ho_1$ and $ho_2$.
These points map (backwards in time) into the sequences:
$ho'_1, \ldots$ and $ho'_2, \ldots$, respectively,
which constitute the primary homoclinic orbits defining two relevant homoclinic areas,
$S_{{ho}_1}$ and $S_{{ho}_2}$, in phase space.
Now, comparing this with Fig.~\ref{fig:2}(a) we see that the density maxima in
the QSOS appear located at the two primary homoclinic points $ho_1$ and $ho_2$.
If we continue examining Fig.~\ref{fig:2}, it is observed in (b) that
only the maximum on $ho_1$ exists, presenting the QSOS a minimum
of density on the other primary homoclinic point, $ho_2$.
For case (d) the situation is reversed, appearing a minimum on $ho_1$ and a
maximum on $ho_2$.
Finally, we find in (c) a case with two minima at $ho_1$ and $ho_2$, respectively.

This behavior can be explained by taking into account that, associated to
the homoclinic orbits, there are fluxes of quantum probability, which follow
the corresponding circuits in phase space.
Each of these homoclinic excursions can be thought of as a channel for this flux.
In these excursions, some part of the quantum probability leaves the vicinity of
the PO and returns later to it with some phase, shifted with respect to the phase
accumulated by the flux along the PO circuit.
When this phase shift is a multiple of $2\pi$, the two probability fluxes interfere
constructively.
In this case, the channel can be considered as a bright channel for the process,
while in the case of a destructive interference the channel will be dark.
The location of these channels is best ascertained by examination of the QSOS,
where the density at the intersection with the homoclinic orbits is monitored.
When one channel (homoclinic orbit) is bright an accumulation of density on the
corresponding homoclinic points will be visible, while for the dark channels
the density at those points is minimum.
Furthermore, the phase shifted by an homoclinic excursion consists of
two different contributions.
One, of dynamical origin, is given by the area enclosed by the homoclinic circuit,
while the other is of topological nature.
Correspondingly, a constructive interference is governed by the following quantization
rule derived in Ref.~\onlinecite{homoclinic}
\begin{equation}
  k S_{ho} = 2 \pi (n_{ho}+\nu_{ho}/4),
  \label{eq:homo}
\end{equation}
where $\nu_{{ho}_1}=-1$ and $\nu_{{ho}_2}=0$ in the case we are considering here.
In this way, the accumulations or lack of probability density observed in
the plots of Fig.~\ref{fig:2} can be explained as the result of the quantization
and/or anti--quantization of the primary homoclinic areas, $S_{{ho}_1}$ or $S_{{ho}_2}$.
This is numerically illustrated in Table \ref{table:I}, where the associated
``homoclinic'' quantum numbers, $n_{{ho}_1}$ and $n_{{ho}_2}$,
computed at the quantized wavelength values of the horizontal PO,
$k_{\rm BS}$, (taken as approximations to the true $k_{{ho}_{1,2}}$ values)
are given.
Moreover, the four cases selected correspond to situations in which $n_{{ho}_{1,2}}$ are,
to very good accuracy, either integer (quantization) or half--integer (anti--quantization)
numbers.
%
\begin{figure}[t]
 \includegraphics[width=7cm]{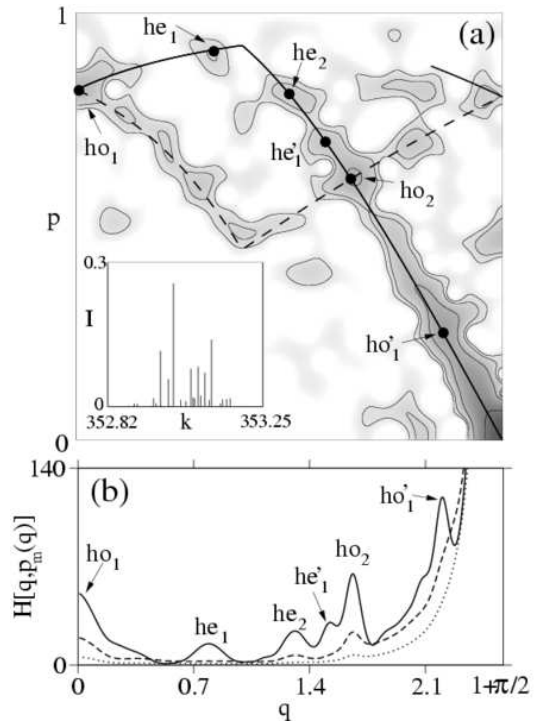}
 \caption{(a) Husimi based quantum surface of section (QSOS) for the horizontal
 scar function with $n=224$ and $T=3.3 t_E$.
 The inset shows the intensities in the basis set of the stadium eigenfunctions.
 (b) Profile of the QSOS along the unstable manifold (full line).
 The results for $T=0.9 t_E$ (dotted line) and $T=1.2 t_E$ (dashed line)
 are also included.}
 \label{fig:4}
\end{figure}

Notice, that in the previous discussion only the two primary homoclinic orbits,
defining the shortest possible homoclinic circuits, have been used.
The question now arises as to what is the influence of other homoclinic orbits,
performing longer and more complicated excursions into phase space regions where
the dynamics are influenced by other POs \cite{Tomsovic}.
These secondary orbits also define phase space circuits that can be quantized
in a way similar to (\ref{eq:homo}).
To investigate their role in the issue addressed in this Letter, we show in
Fig.~\ref{fig:4}(a) the QSOS for the horizontal scar function with $n_H=224$,
computed for $T=3.3 t_E$.
In this case, both primary homoclinic areas are quantized,
with values of $n_{ho_1}=189.01$ and $n_{ho_2}=168.07$.
As can be seen, a lot of fine structure has developed here and several density
accumulations appear along the manifolds.
To analyze more easily the resulting structure, we have plotted in (b) the QSOS profile,
$H[q,p_m(q)]$, along the unstable manifold of the horizontal PO.
Three values of the evolution time have been considered, corresponding to
$T/t_E=0.9,\; 1.2,\; 3.3$, respectively.
For the shortest value (dotted line), only the maxima on the primary homoclinic
points, $ho_1$ and $ho_2$, are (barely) visible.
This result is in agreement with our previous conclusion about the results of
Fig.~\ref{fig:2}(a).
As the evolution time increases, and we move to $T=1.2 t_E$ (dashed line),
the structure is better resolved.
As a consequence, the two previous maxima get more prominent,
and also a new peak centered on $he_2$ appears.
Finally, at the longest time considered, $T=3.3 t_E$, all previous maxima get
better defined, two extra maxima appear on $he_1$ and $he'_1$,
and even successive mappings of previous points, namely $ho'_1$, are resolved
within the big peak at $q_m=1+\pi/2$ [corresponding to (A) in Fig.~\ref{fig:3}].
Following our previous discussion, the existence of these new peaks could
be explained as a consequence of the quantization of the phase space circuit
defined by secondary homoclinic orbits.
Rather, we prefer to use here an argument based on heteroclinic orbits.
The advantage being that in this way a family of homoclinic orbits
leaving the original PO, making an integer number of turns around a
new visited short PO, and returning again to the vicinity of the original PO,
can be substituted by a single heteroclinic circuit \cite{Vergini3,heteroclinic}.
This circuit consists of an heteroclinic orbit that populates the new PO by
transporting probability from the original PO,
and another heteroclinic orbit transporting probability in the opposite direction.
In this way, long term dynamics are dealt with more efficiently,
thus avoiding the serious computational problem associated to the
exponential proliferation of homoclinic orbits.
These heteroclinic circuits are identified, on a surface of section, by
four consecutive pieces of manifold, defining an invariant (heteroclinic)
area in phase space.
For instance, the heteroclinic circuit between the horizontal PO and the
``flipped L''--shaped (rendering an hexagon in the full version of the stadium) PO
(see Fig.~\ref{fig:3} center) is given by:
the piece of unstable manifold from (A) to $he'_1$,
the piece of stable manifold from $he'_1$ to the (B),
the piece of unstable manifold from (B) to the heteroclinic point ${he}^*$
(belonging to the heteroclinic orbit transporting probability towards the
horizontal PO again),
and finally, the piece of stable manifold from ${he}^*$ to (A).
The transverse area enclosed by the circuit (shaded in the figure) is $S_{{he}_1}$.
We emphasize that by evolving this circuit forward and backward in time,
an heteroclinic tube characterized by $S_{{he}_1}$ is defined.
Having this in mind, the point $he'_1$ and its mapping $he_1$ are assigned to
the same heteroclinic channel, which will be bright when the following quantization
condition is fulfilled,
\begin{equation}
  k S_{{he}_1} =2 \pi n_{{he}_1}.
  \label{eq:4}
\end{equation}
This condition, derived from the fact that the overlap between the scar function
for $T>2t_E$ and the flipped--L tube function is proportional to
$\cos(k S_{{he}_1}/2)$ \cite{Vergini3}, guarantees that the region around the
flipped--L PO is effectively populated  through the heteroclinic channel
described above.
In our case, $n_{{he}_1} \simeq 19.00$, which is extremely close to an integer value,
thus explaining the existence of the corresponding maxima in the QSOS.
Another maximum appear at $he_2$ in Fig.~\ref{fig:4}.
It can be explained through the heteroclinic circuit associated to the
``triangle'' (``kite'' in the full stadium) PO and related area $S_{{he}_2}$
shown in the left part of Fig.~\ref{fig:3}.
In this case the computed value for the ``heteroclinic'' quantum number is
$n_{{he}_2} \simeq 5.98$, again extremely close to an integer value.

A final point worth discussing concerns the spectral localization of the scar
functions (\ref{eq:scar}), that have been used throughout this Letter.
For this purpose, an inset with the squared modulus of the coefficients
projecting the corresponding scar function on the stadium eigenfunctions
has been included in Fig.~\ref{fig:4}.
As can be seen, it essentially consists of only 7 contributions,
a terribly small number if one takes into account that a generic plane wave
(usually taken as optimal basis functions for the evaluation of billiard
eigenfunctions) with the same $k$ requires at least 300 eigenfunctions
for a correct description.
This clearly indicates that the scarring effect by homoclinic and heteroclinic
orbits that we have unveiled here really opens the door for a systematic
construction of the eigenfunctions of classically chaotic systems.

\begin{acknowledgments}
This work was supported by MCyT, Comunidad de Madrid, UPM, and AECI (Spain)
under contracts BQU2003--8212, P--TIC--191-0505, AL05--PID--18 and (A/4574/05
and A/3752/05), respectively, UAM--Grupo Santander, and CONICET (Argentina).
\end{acknowledgments}
%

%
\end{document}